\documentclass[12pt]{article}
\usepackage{amsmath,epsfig,graphicx,amssymb}
\begin{document}
\begin{center}
{\large \bf  Bounding the Magnetic and Electric Dipole Moments of $\nu_\tau$
             from the Process $e^{+}e^{-}\rightarrow \nu \bar \nu \gamma$ in $E_6$
             Superstring Models}
\end{center}

\baselineskip 20.0pt

\vspace{2mm}

\begin{center}
{\small {A. Guti\'errez-Rodr\'{\i}guez $^{1, a}$, M. A. Hern\'andez-Ru\'{\i}z $^{2, a}$,
         B. Jayme-Vald\'es $^1$  and M. A. P\'erez $^3$ }}  \\[2mm]
{\small {\it (1)  Facultad de F\'{\i}sica, Universidad Aut\'onoma de Zacatecas\\
            Apartado Postal C-580, 98060 Zacatecas, Zacatecas M\'exico.\\
        (a) Cuerpo Acad\'emico de Part\'{\i}culas, Campos y Astrof\'{\i}sica. \\[3mm]
}}

{\small {\it (2) Facultad de Ciencias Qu\'{\i}micas, Universidad Aut\'onoma de Zacatecas\\
             Apartado Postal 585, 98060 Zacatecas, Zacatecas M\'exico.  \\[2mm]
}}

{\small {\it (3) Departamento de F\'{\i}sica, CINVESTAV.\\
             Apartado Postal 14-740, 07000 M\'exico D. F., M\'exico. \\[2mm]
}}


\date{\today}

\vspace{1cm}

\begin{minipage}{5in}
\centerline{{\sc Abstract}}
\medskip
\baselineskip 22.1pt plus 0.2pt minus 0.1pt

{\small
We obtain bounds on the anomalous magnetic and electric
dipole moments of the tau-neutrino through the reaction
$e^{+}e^{-}\rightarrow \nu \bar \nu \gamma$ at the $Z_1$-pole
in the framework of a Left-Right symmetric model and a
class of $E_6$ inspired models with an additional neutral vector
boson $Z_\theta$. We use the data collected by the L3 Collaboration
at LEP. For the parameters of the $E_6$ model we consider the mixing
angle $\theta_{E_6}=37.8^o$ and $M_{Z_\theta}=7M_{Z_1}$. We find that
our bounds are of the same order of magnitude as those obtained in other
extensions of the Standard Model.}
\end{minipage}
\end{center}

\vspace{3mm}

\noindent{{PACS:} 14.60.St, 13.40.Em, 12.15.Mm, 14.60.Fg.\\

\newpage

\section{Introduction}
In the Standard Model (SM) \cite{S.L.Glashow} extended to contain right-handed
neutrinos, the neutrino magnetic moment induced by radiative corrections is
unobservably small, $\mu_\nu \sim 3\times 10^{-19}(m_\nu/1 \hspace{1mm} eV)$ \cite{Mohapatra}.
Current limits on these magnetic moments are several orders of magnitude larger,
so that a magnetic moment close to these limits would indicate a window for
probing effects induced by new physics beyond the SM \cite{Fukugita}. Similarly,
a neutrino electric dipole moment will point also to new physics and they will
be of relevance in astrophysics and cosmology, as well as terrestial neutrino
experiments \cite{Cisneros}.

The existence of a heavy neutral ($Z'$) vector boson is a feature
of many extensions of the standard model. In particular, one (or
more) additional $U(1)'$ gauge factor provides one of the simplest
extensions of the SM. Additional $Z'$ gauge bosons appear in Grand
Unified Theories (GUT's) \cite{Robinett}, Superstring Theories
\cite{Green}, Left-Right Symmetric Models (LRSM)
\cite{Mohapatra,G.Senjanovic,G.Senjanovic1} and in other models
such as models of composite gauge bosons \cite{Schrempp}. The
largest set of extended gauge theories are those which are based
on GUT's. Popular examples are the groups $SO(10)$ and $E_6$.
Generically, additional $Z$-bosons originating from $E_6$ grand
unified theories are conveniently labeled in terms of the chain:
$E_6 \to SO(10)\times U(1)_\psi \to SU(5)\times U(1)_\chi \times
U(1)_\psi \to SM \times U(1)_{\theta_{E_6}}$ where
$U(1)_{\theta_{E_6}}$ remains unbroken at low energies. Detailed
discussions on GUTS can be found in the literature
\cite{Robinett,Green}.

T. M. Gould and I. Z. Rothstein \cite{T.M.Gould} reported in 1994
a bound on $\mu_{\nu_\tau}$ obtained through the analysis of the
process $e^{+}e^{-} \rightarrow \nu \bar \nu \gamma$, near the
$Z_1$-resonance, with a massive neutrino and the SM
$Z_1e^{+}e^{-}$ and $Z_1\nu \bar \nu$  couplings. In this process,
the dependence of $\mu_{\nu_\tau}$ and $d_{\nu_\tau}$ comes from
the radiation of the photon by the neutrino and antineutrino in
the final state. The Feynman diagrams which give the most
important contribution to the cross-section are shown in Fig. 1.
We stress here the importance of the final state radiation near
the $Z_1$ pole of a very energetic photon as compared to the
conventional Bremsstrahlung. The study of the same process in the
framework of a LRSM was reported recently \cite{Alexgu}. It was
found that the L3 data obtained at LEP \cite{L3} induce bounds on
$\mu_{\nu_\tau}$ and $d_{\nu_\tau}$ which are almost independent
of the mixing angle between $Z_1$ and the new heavy $Z_2$ gauge
boson predicted in LRSM and $M_{Z_2}$.

Our aim in the present paper is to analyze the reaction
$e^{+}e^{-}\rightarrow \nu \bar\nu \gamma$ at the $Z_1$ boson
resonance and in the framework of a LRSM and in a class of $E_6$
inspired models with a light additional neutral vector boson
$Z_\theta$ and we attribute an anomalous magnetic moment (MM) and
an electric dipole moment (EDM) to a massive tau neutrino.
Processes measured near the resonance serve to set limits on the
tau neutrino MM and EDM. In this paper, we take advantage of this
fact to set bounds for $\mu_{\nu_{\tau}}$ and $d_{\nu_{\tau}}$ for
various values of the mixing angle $\phi$ of the LRSM and for
$\theta_{E_6}=37.8^o$ and $M_{Z_\theta}=7M_{Z_1}$, the parameters
of a class of $E_6$ inspired models, according to Ref. \cite{Aytekin}.

The L3 Collaboration evaluated the selection efficiency using detector-simulated
$e^{+}e^{-}\rightarrow \nu \bar \nu \gamma (\gamma)$ events, random trigger
events, and large-angle $e^{+}e^{-}\rightarrow e^+e^-$ events. A total of 14 events
were found by the selection. The distributions of the photon energy and the
cosine of its polar angle are consistent with SM predictions.

This paper is organized as follows: In Sec. 2 we describe the neutral current
couplings in $E_6$. In Sect. 3 we present the calculation of the process
$e^{+}e^{-}\rightarrow \nu \bar\nu \gamma$ with an extra $Z_\theta$ boson.
Finally, we present our results and conclusions in Sect. 4.

\vspace{3mm}

\section{Neutral Current Couplings in $E_6$}

\vspace{3mm}

In this section we describe the neutral current couplings involved in the class
of $E_6$ inspired models we are interested in. Let us consider the following
breakdown pattern in $E_6$:

\begin{equation}
E_6 \to SO(10)\times U(1)_\psi \to SU(5)\times U(1)_\chi \times U(1)_\psi \to SM \times U(1)_{\theta_{E_6}},
\end{equation}

\noindent where the $SU(3)_C \times SU(2)_L \times U(1)_{Y_W}$ groups of the
standard model are embeded in $SU(5)$ of $SO(10)$. The couplings of the fermions
to the standard model $Z_1$ are given, as usual, by

\begin{equation}
Q=I_{3L}-Q_{em}\sin^2\theta_W,
\end{equation}

\noindent while the couplings to the $Z_\theta$ are given by linear combinations
of the $U(1)_\chi$ and $U(1)_\psi$ charges \cite{London,Capstick}:

\begin{eqnarray}
Q'&=&Q_\chi \cos\theta_{E_6}+Q_\psi \sin\theta_{E_6}, \nonumber\\
Q''&=&-Q_\chi \sin\theta_{E_6}+Q_\psi \cos\theta_{E_6},
\end{eqnarray}

\noindent where the operators $Q_\psi$ and $Q_\chi$ are orthogonal to those
of $Q_{em}$ and that of the standard model $Z_1$ and $\theta_{E_6}$ is the
$Q_\chi-Q_\psi$ mixing angle in $E_6$.

With the extra $Z_\theta$ neutral vector boson the neutral current Lagrangian is \cite{Leike,Alam}

\begin{equation}
{-\cal L}_{NC}=eA^\mu J_{em\mu}+g_1Z^\mu_1 J_{Z_{1}\mu}+g_2Z^\mu_\theta J_{Z_{\theta}\mu},
\end{equation}

\noindent where $J_{em \mu}$, $J_{Z_{1}\mu}$ and $J_{Z_{\theta}\mu}$ are the electromagnetic
current, the $Z_1$ current of the standard model and the $J_{Z_\theta\mu}$ current
of the new boson, respectively and are given by

\begin{eqnarray}
J_{Z_1 \mu}&=&\sum_f\bar f\gamma_\mu(C^1_V+C^1_A\gamma_5)f, \nonumber\\
J_{Z_\theta \mu}&=&\sum_f\bar f\gamma_\mu(C'_V+C'_A\gamma_5)f,
\end{eqnarray}

\noindent where $f$ represents fermions, while

\begin{eqnarray}
g_1&=&(g^2+g'^2)^{1/2}= \frac{e}{2\sin\theta_W\cos\theta_W}, \nonumber\\
g_2&=&g_{\theta},\\
C^{1e}_V&=&-\frac{1}{2}+2\sin^2 {\theta_W}, \hspace{1cm} C^{1e}_A=\frac{1}{2},\nonumber\\
C^{1\nu}_V&=&\frac{1}{2},\hspace*{3.4cm} C^{1\nu}_A=\frac{1}{2},\\
C^{'e}_V&=&z^{1/2}(\frac{\cos\theta_{E_6}}{\sqrt{6}}+\frac{\sin\theta_{E_6}}{\sqrt{10}}),\hspace{1.5cm} C^{'e}_A=2z^{1/2}\frac{\sin\theta_{E_6}}{\sqrt{10}}, \nonumber\\
C^{'\nu}_V&=&z^{1/2}(-\frac{\cos\theta_{E_6}}{\sqrt{6}}+\frac{3\sin\theta_{E_6}}{\sqrt{10}}),\hspace*{9mm} C^{'\nu}_A=z^{1/2}(-\frac{\cos\theta_{E_6}}{\sqrt{6}}+\frac{3\sin\theta_{E_6}}{\sqrt{10}}),
\end{eqnarray}

\noindent with

\begin{equation}
z=(\frac{g^2_\theta}{g^2+g'^2})(\frac{M_{Z_1}}{M_{Z_\theta}})^2,
\end{equation}

\noindent a parameter that depends on the coupling constant $g_\theta$ and
$M_{Z_\theta}$.

The class of $E_6$ models we shall be interested in arise with the following
specific values for the mixing angle $\theta_{E_6}$ \cite{London}:

\begin{eqnarray}
\theta_{E_6}&=&0^o, \hspace{6mm} Z_{\theta_{E_6}}\to Z_\psi,\nonumber\\
\theta_{E_6}&=&37.8^o, \hspace{6mm} Z_{\theta_{E_6}}\to Z', \nonumber\\
\theta_{E_6}&=&90^o, \hspace{6mm} Z_{\theta_{E_6}}\to Z_\chi,\\
\theta_{E_6}&=&127.8^o, \hspace{6mm} Z_{\theta_{E_6}}\to Z_I,\nonumber
\end{eqnarray}

\noindent where $Z_\psi$ is the extra neutral gauge boson arising in
$E_6 \to SO(10)\times U(1)_\psi$,
$Z'$ corresponds to the respective neutral gauge boson obtained if $E_6$ is
broken down to a rank-5 group, $Z_\chi$ is the neutral gauge boson involved in
$SO(10) \to SU(5)\times U(1)_\chi$,
and $Z_I$ is the neutral gauge boson associated to the breaking of $E_6$ via
a non-Abelian discrete symmetry to a rank-5 group \cite{London}.

\vspace{3mm}

\section{The Total Cross Section}

\vspace{3mm}

We will take advantage of our previous work on the LRSM and we will calculate
the total cross section for the
reaction $e^{+}e^{-}\rightarrow \nu \bar\nu \gamma$ using the
transition amplitudes given in Eqs. (14) and (15) of Ref.
\cite{Alexgu} for the LRSM for diagrams 1 and 2 of Fig. 1. For the
contribution coming from diagrams 3 and 4 of Fig. 1, we use Eqs. (5) and (8)
given in section 2 for the $E_6$ model. The respective transition
amplitudes are thus given by

\begin{eqnarray}
{\cal M}_{1}&=&\frac{-g^{2}}{8\cos^{2}\theta_{W}(l^{2}-m^{2}_{\nu})}[\bar u(p_{3})\Gamma^{\alpha}(\ell\llap{/}+m_{\nu})\gamma^{\beta}(a-b\gamma_{5})v(p_{4})]\nonumber\\
         &&\frac{(g_{\alpha\beta}-p_{\alpha}p_{\beta}/M^{2}_{Z_1})}{[(p_{1}+p_{2})^{2}-M^{2}_{Z_1}-i\Gamma^{2}_{Z_1}]}[\bar u(p_{2})\gamma^{\alpha}(aC_V^{1e}-bC_A^{1e}\gamma_{5})v(p_{1})]\epsilon^{\lambda}_{\alpha},
\end{eqnarray}


\begin{eqnarray}
{\cal M}_{2}&=&\frac{-g^{2}}{8\cos^{2}\theta_{W}(k^{2}-m^{2}_{\nu})}[\bar u(p_{3})\gamma^{\beta}(a-b\gamma_{5})(k\llap{/}+m_{\nu})\Gamma^{\alpha}   v(p_{4})]\nonumber\\
         &&\frac{(g_{\alpha\beta}-p_{\alpha}p_{\beta}/M^{2}_{Z_1})}{[(p_{1}+p_{2})^{2}-M^{2}_{Z_1}-i\Gamma^{2}_{Z_1}]}[\bar u(p_{2})\gamma^{\alpha}(aC_V^{1e}-bC_A^{1e}\gamma_{5})v(p_{1})]\epsilon^{\lambda}_{\alpha},
\end{eqnarray}

\noindent and for $M'_1$ and $M'_2$

\begin{eqnarray}
M'_1&=&M_1(a \to C_V', b \to C_A', M_{Z_1} \to M_{Z_\theta}),\\
M'_2&=&M_2(a \to C_V', b \to C_A', M_{Z_1} \to M_{Z_\theta}),
\end{eqnarray}

\noindent where

\begin{equation}
\Gamma^{\alpha}=eF_{1}(q^{2})\gamma^{\alpha}+\frac{ie}{2m_{\nu}}F_{2}(q^{2})\sigma^{\alpha
\mu}q_{\mu}+ eF_{3}(q^{2})\gamma_5\sigma^{\alpha \mu}q_{\mu},
\end{equation}

\noindent is the neutrino electromagnetic vertex, $e$ is the
charge of the electron, $q^\mu $ is the photon momentum and
$F_{1,2,3}(q^2)$ are the electromagnetic form factors of the
neutrino, corresponding to charge radius, MM and EDM,
respectively, at $q^2=0$ \cite{Escribano,Vogel}, while
$\epsilon^{\lambda}_{\alpha}$ is the polarization vector of the
photon. $l$ ($k$) stands for the momentum of the virtual neutrino
(antineutrino), and the coupling constants $a$ and $b$ are given
in the Eq. (15) of the Ref. \cite{Alexgu}, while $C'_V$ and $C'_A$
are given above in the Eq. (8).

Using the same notation as in the Ref. \cite{T.M.Gould}, we find
that the MM, EDM, the mixing angle $\phi$ of the LRSM as well as
the mixing angle $\theta_{E_6}$ and the mass of the additional
neutral vector boson $M_{Z_\theta}$ of the $E_6$
model give a contribution to the differential cross section for
the process $e^{+}e^{-}\rightarrow \nu \bar \nu \gamma$ of the
form:

\begin{eqnarray}
\frac{d\sigma}{E_{\gamma}dE_{\gamma}d\cos\theta_{\gamma}}
&=&\frac{\alpha^{2}}{192\pi}\;
[\mu^{2}_{\nu_\tau}+d^{2}_{\nu_{\tau}}]\;
[{\cal C}[\phi, x_{W}]\;
{\cal F}[\phi, s, E_{\gamma}, \cos\theta_{\gamma}]\nonumber \\
&+&{\cal C}_1[\theta_{E_6}, M_{Z_\theta}, x_{W}]\;
{\cal F}_1[M_{Z_\theta}, s, E_{\gamma}, \cos\theta_{\gamma}]
+8f[M_{Z_1}, \Gamma_{Z_1}, M_{Z_\theta}, \Gamma_{Z_\theta}, s, E_\gamma, \cos\theta_\gamma]\nonumber\\
&\cdot&\{ {\cal C}_2[\phi, \theta_{E_6}, M_{Z_\theta}, x_{W}]\;
{\cal F}_2[s]
+ {\cal C}_3[\phi, \theta_{E_6}, M_{Z_\theta}, x_{W}]\;
{\cal F}_3[s, E_{\gamma}]
+ {\cal C}_4[\theta_{E_6}, M_{Z_\theta}]\;
{\cal F}_4[E_{\gamma}]\nonumber\\
&+& {\cal C}_5[\phi, \theta_{E_6}, M_{Z_\theta}, x_{W}]\;
{\cal F}_5[s, E_{\gamma},\cos\theta_{\gamma}]
+ {\cal C}_6[\phi, \theta_{E_6}, M_{Z_\theta}, x_{W}]\;
{\cal F}_6[s, E_{\gamma},\cos\theta_{\gamma}]\}]
\end{eqnarray}

\noindent where $E_{\gamma}$, $\cos\theta_{\gamma}$ are the energy
and scattering angle of the photon.\\

The kinematics is contained in the functions

\begin{eqnarray}
{\cal F}[\phi, s, E_{\gamma}, \cos\theta_{\gamma}]
&\equiv &\frac{(a^{2}+b^{2}) (s-2\sqrt{s}E_{\gamma})+b^{2}E^{2}_{\gamma}\sin^{2}\theta_{\gamma}}
{(s-M_{Z_1}^2)^2+M^{2}_{Z_1}\Gamma^{2}_{Z_1}},\nonumber\\
{\cal F}_1[M_{Z_\theta}, s, E_{\gamma}, \cos\theta_{\gamma}]
&\equiv & \frac{6(s-2\sqrt{s}E_{\gamma})+3E^{2}_{\gamma}\sin^{2}\theta_{\gamma}}
{(s-M_{Z_\theta}^2)^2+M^{2}_{Z_\theta}\Gamma^{2}_{Z_\theta}},\nonumber\\
{\cal F}_2[s]
&\equiv & 4\sqrt{s},\\
{\cal F}_3[s, E_{\gamma}]
&\equiv & 2\sqrt{s}E_{\gamma},\nonumber\\
{\cal F}_4[E_{\gamma}]
&\equiv & \sqrt{15}E_{\gamma},\nonumber\\
{\cal F}_5[s, E_{\gamma}, \cos\theta_{\gamma}]
&\equiv & -(s+\frac{1}{2}E^{2}_{\gamma}\sin^{2}\theta_{\gamma}),\nonumber\\
{\cal F}_6[s, E_{\gamma}, \cos\theta_{\gamma}]
&\equiv & (s-E_{\gamma}^{2}+\frac{1}{2}E^{2}_{\gamma}\sin^{2}\theta_{\gamma}),\nonumber
\end{eqnarray}

\begin{eqnarray}
&&f(s, M_{Z_1}, \Gamma_{Z_1}, M_{Z_\theta}, \Gamma_{Z_\theta})\nonumber \\
&\equiv &
\frac{-2[(s-M^2_{Z_{1}})(s-M^2_{Z_{\theta}})+M_{Z_{1}}\Gamma_{Z_{1}}M_{Z_{\theta}}\Gamma_{Z_{\theta}}]}
{[(s-M^2_{Z_{1}})(s-M^2_{Z_{\theta}})+M_{Z_{1}}\Gamma_{Z_{1}}M_{Z_{\theta}}\Gamma_{Z_{\theta}}]^2
+[(s-M^2_{Z_{\theta}})M_{Z_{1}}\Gamma_{Z_{1}}-(s-M^2_{Z_{1}})M_{Z_{\theta}}\Gamma_{Z_{\theta}}]^2}.
\end{eqnarray}

The coefficients ${\cal C}$, ${\cal C}_1$,..., ${\cal C}_6$ are given by

\begin{eqnarray}
{\cal C}[\phi, x_{W}]
&\equiv &\frac{[\frac{1}{2}(a^{2}+b^{2})-4a^{2}x_{W}
+ 8a^{2}x^{2}_{W}]}{x^{2}_{W}(1 - x_{W})^{2}},\nonumber\\
{\cal C}_1[\theta_{E_6}, M_{Z_\theta}, x_{W}]
&\equiv &\frac{(C_V^{2}+ C_A^{2})}{x^{2}_{W}(1 - x_{W})^{2}},\nonumber\\
{\cal C}_2[\phi, \theta_{E_6}, M_{Z_\theta}, x_{W}]
&\equiv&\frac{[2ax_{W}-\frac{1}{2}(a+ b)][C_V^{'e}C_A^{'e}-(C_A^{'e})^{2}]}{x^{2}_{W}(1-x_{W})^{2}},\nonumber\\
{\cal C}_3[\phi, \theta_{E_6}, M_{Z_\theta}, x_{W}]
&\equiv&\frac{[2ax_{W}+\frac{1}{2}(a+ b)][3(C_A^{'e})^{2}-(C_V^{'e}-C_A^{'e})^{2}]}{x^{2}_{W}(1-x_{W})^{2}}\\
{\cal C}_4[\theta_{E_6}, M_{Z_\theta}, x_{W}]
&\equiv&\frac{[3(C_A^{'e})^{2}+(C_V^{'e}-C_A^{'e})^{2}]}{x^{2}_{W}(1-x_{W})^{2}},\nonumber\\
{\cal C}_5[\phi, \theta_{E_6}, M_{Z_\theta}, x_{W}]
&\equiv&\frac{3(C_A^{'e})^{2}[2ax_{W}+\frac{1}{2}(a+ b+1)]+
[4ax_{W}-(a+ b)][C_V^{'e}C_A^{'e}-(C_A^{'e})^{2}]}{x^{2}_{W}(1-x_{W})^{2}},\nonumber\\
{\cal C}_6[\phi, \theta_{E_6}, M_{Z_\theta}, x_{W}]
&\equiv&\frac{[2ax_{W}-\frac{1}{2}a](C_V^{'e}-C_A^{'e})^{2}}{x^{2}_{W}(1-x_{W})^{2}},\nonumber
\end{eqnarray}

\noindent where $C_V$ and $C_A$ are now given by:

\begin{eqnarray}
C_V&=&z(\frac{\cos\theta_{E_6}}{\sqrt{6}}+\frac{\sin\theta_{E_6}}{\sqrt{10}})(-\frac{\cos\theta_{E_6}}{\sqrt{6}}+\frac{3\sin\theta_{E_6}}{\sqrt{10}}),\nonumber\\
C_A&=&2z\frac{\sin\theta_{E_6}}{\sqrt{10}}(-\frac{\cos\theta_{E_6}}{\sqrt{6}}+\frac{3\sin\theta_{E_6}}{\sqrt{10}}),
\end{eqnarray}

\noindent with $x_{W}\equiv \sin^{2}\theta_{W}$.\\

In the above expressions, the function ${\cal F}$ includes the
contribution coming from the exchange of the SM/LRSM $Z_1$ gauge
boson, ${\cal F}_1$ includes the contribution arising from the
exchange of the heavy gauge boson $Z_\theta$, while the function
$f$ contains the interference coming from both exchanges. Taking
the limit when $M_{Z_\theta}\to \infty$ and the mixing angle
$\phi=0$, the expressions for $C_V^{'e,\nu}$ and $C_A^{'e,\nu}$
reduce to $C_V^{'e,\nu}=C_A^{'e,\nu}=C_V=C_A=0$ and, the Eq. (16)
reduces to the expression (3) given in Ref. \cite{T.M.Gould} for
the SM. On the other hand, taking the limit when $M_{Z_\theta}\to
\infty$ the Eq. (16) reduces to the expressions (25) given in Ref.
\cite{Alexgu} for the LRSM. Finally, if the mixing angle is taken
as $\phi=0$ the Eq. (16) reduce to the expression (24) given in
Ref. \cite{Aytekin} for the $E_6$ model.

\section{Results and Conclusions}

\vspace{3mm}

In order to evaluate the integral of the total cross section as a function of
the parameters of the LRSM-$E_6$ models, that is to say, $\phi$, $M_{Z_\theta}$
and the mixing angle $\theta_{E_6}$, we require cuts on the photon angle and
energy to avoid divergences when the integral is evaluated at the important
intervals of each experiment. We integrate over $\theta_\gamma$ from $44.5^o$
to $135.5^o$ and $E_\gamma$ from 15 $GeV$ to 100 $GeV$ for various fixed values
of the mixing angle $\phi=-0.009, -0.004, 0, 0.004$ and for $\theta_{E_6} = 37.8^o$
(which corresponds to $Z_\theta \to Z'$) and $M_{Z_\theta}= 7M_{Z_1}$ according
to the Ref. \cite{Aytekin}. Using the following numerical values: $\sin^2\theta_W=0.2314$,
$M_{Z_1}=91.18$ $GeV$, $\Gamma_{Z_\theta}=\Gamma_{Z_1}=2.49$ $GeV$ and
$z=(\frac{3}{5}\sin^2\theta_W)(\frac{M_{Z_1}}{M_{Z_\theta}})^2$
we obtain the cross section $\sigma=\sigma(\phi, \theta_{E_6}, M_{Z_\theta}, \mu_{\nu_\tau},d_{\nu_\tau})$.

For the mixing angle $\phi$ between $Z_1$ and $Z_2$ of the LRSM, we use the
reported data of Maya {\it et al.} \cite{M.Maya}:

\begin{equation}
-9\times 10^{-3}\leq \phi \leq 4\times 10^{-3},
\end{equation}

\noindent with a $90 \%$ C. L.

Since we have calculated the cross-section at the $Z_1$ pole, {\it i.e.} at
$s=M^2_{Z_1}$, the value of $\sin^2\theta_W$ is not affected by the $Z_\theta$
physics \cite{Langacker1,Demir}. Variation of the $\Gamma_{Z_\theta}$ is taken
in the range from 0.15 to 2.0 times $\Gamma_{Z_1}$ in the results of the CDF Collaboration
\cite{CDF}. So we take  $\Gamma_{Z_\theta}=\Gamma_{Z_1}$ as a special case of
this variation.

As was discussed in Ref. \cite{T.M.Gould}, $N\approx\sigma(\phi, \theta_{E_6}, M_{Z_\theta}, \mu_{\nu_\tau}, d_{\nu_\tau}){\cal L}$.
Using the Poisson statistic \cite{L3,Barnett}, we require that $N\approx\sigma(\phi, \theta_{E_6}, M_{Z_\theta}, \mu_{\nu_\tau}, d_{\nu_\tau}){\cal L}$
be less than 14, with ${\cal L}= 137$ $pb^{-1}$, according to the data
reported by the L3 Collaboration Ref. \cite{L3} and references therein.
Taking this into consideration, we can get a bound for the tau neutrino magnetic
moment as a function of $\phi$, $\theta_{E_6}$ and $M_{Z_\theta}$ with
$d_{\nu_\tau}=0$. The values obtained for this bound for several values of
$\phi$ with $\theta_{E_6}=37.8^o$ and $M_{Z_\theta}=7M_{Z_1}$ are included
in Table 1.\\

\begin{center}
\begin{tabular}{|c|c|c|}\hline
$\phi$&$\mu_{\nu_\tau} (10^{-6}\mu_B)$&$d_{\nu_\tau}(10^{-17}e \mbox{cm})$\\
\hline \hline
-0.009&3.37&6.50\\
\hline
-0.004&3.33&6.43\\
\hline
0&3.31&6.40\\
\hline
0.004&3.30&6.33\\
\hline
\end{tabular}
\end{center}

\begin{center}
Table 1. Bounds on the $\mu_{\nu_\tau}$ magnetic moment and
$d_{\nu_\tau}$ electric dipole moment for different values of the
mixing angle $\phi$ with $\theta_{E_6}=37.8^o$ and $M_{Z_\theta}=7M_{Z_1}$.
We have applied the cuts used by L3 for the photon angle and energy.
\end{center}

\vspace{3mm}

The results obtained in Table 1 are in agreement with the literature
\cite{T.M.Gould,L3,Escribano,DELPHI,J.M.Hernandez,F.Larios,H.Grotch}.
However, if the photon angle and energy are $0\leq \theta_\gamma
\leq \pi$ and $15\; GeV\leq E_\gamma \leq 100 \; GeV$ with
$\phi=-0.009, -0.004, 0, 0.004$, $\theta_{E_6} = 37.8^o$,
$M_{Z_\theta}= 7M_{Z_1}$, $N=14$ and ${\cal L}=48$ $pb^{-1}$,
we obtained the results given in Table 2.\\

\newpage

\begin{center}
\begin{tabular}{|c|c|c|}\hline
$\phi$&$\mu_{\nu_\tau} (10^{-6}\mu_B)$&$d_{\nu_\tau}(10^{-17}e \mbox{cm})$\\
\hline \hline
-0.009&1.85&3.58\\
\hline
-0.004&1.84&3.55\\
\hline
0&1.83&3.53\\
\hline
0.004&1.82&3.52\\
\hline
\end{tabular}
\end{center}

\begin{center}
Table 2. Bounds on the $\mu_{\nu_\tau}$ magnetic moment and
$d_{\nu_\tau}$ electric dipole moment for different values of the
mixing angle $\phi$ with $\theta_{E_6}=37.8^o$ and $M_{Z_\theta}=7M_{Z_1}$.
In this case, we did not use cuts for the photon angle and energy.
\end{center}

\vspace*{3mm}

The previous analysis and comments can readily be translated to the EDM
of the $\tau$-neutrino with $\mu_{\nu_\tau}=0$. The resulting
bounds for the EDM as a function of $\phi$, $\theta_{E_6}$ and $M_{Z_\theta}$
are shown in Tables 1 and 2.

We plot the total cross section in Fig. 2 as a function of the
mixing angle $\phi$ for the bounds of the magnetic moment given in
Tables 1, and 2 with $\theta_{E_6}=37.8^0$ and
$M_{Z_\theta}=7M_{Z_1}$. We reproduce the Fig. 2 of the Ref.
\cite{Alexgu}.

We have determined bounds on the magnetic moment and the electric
dipole moment of a massive tau neutrino in the framework of a LRSM
and a class of $E_6$ inspired models with a light additional
neutral vector boson, as a function of $\phi$, $M_{Z_\theta}$ and
the mixing angle $\theta_{E_6}$, as shown in Tables 1 and 2.

In a previous paper \cite{Alexgu} we estimated bounds on the anomalous magnetic moment and the
electric dipole moment of the tau neutrino through the process
$e^{+}e^{-}\rightarrow \nu \bar\nu \gamma$ in the context of a LRSM
at the $Z_1$ pole. We found that the bounds are almost independent of the
mixing angle $\phi$ of the model. In the present paper we reproduce these
bounds for $\theta_{E_6}=37.8^o$ and $M_{Z_\theta}=7M_{Z_1}$,
corresponding to the $E_6$ superstring models. In Ref. \cite{Aytekin},
Aydemir {\it et al.} analyzed the same process $e^{+}e^{-}\rightarrow \nu \bar\nu \gamma$
also in $E_6$ models. Their analytical and numerical results for MM are
similar than ours, and we are able to reproduce their limits for $\phi =0$,
$0\leq \theta_\gamma \leq \pi$ and ${\cal L}=48$ $pb^{-1}$. These results are
in agreement with the bounds obtained in previous studies \cite{F.Larios,H.Grotch,Maya},
but are well above the SM effects induced by one-loop diagrams \cite{J.M.Hernandez}.

Other upper limits on the tau neutrino magnetic moment reported in the literature
are $\mu_{\nu_{\tau}} < 3.3 \times 10^{-6} \mu_{B}$  $(90 \hspace{1mm}\% \hspace{1mm} C.L.)$
from a sample of $e^{+}e^{-}$ annihilation events collected with the L3 detector at the $Z_1$
resonance corresponding to an integrated luminosity of $137$ $pb^{-1}$ \cite{L3};
$\mu_{\nu_{\tau}} \leq 2.7 \times 10^{-6} \mu_{B}$ $(95 \hspace{1mm} \% \hspace{1mm} C.L.)$
at $q^2=M^2_{Z_1}$ from measurements of the $Z_1$ invisible width at LEP \cite{Escribano};
$\mu_{\nu_{\tau}} \leq 2.62 \times 10^{-6}$ in the effective Lagrangian approach
at the $Z_1$ pole \cite{Maya};
$\mu_{\nu_{\tau}} < 1.83 \times 10^{-6} \mu_{B}$ $(90 \hspace{1mm} \% \hspace{1mm}C.L.)$
from the analysis of $e^{+}e^{-}\rightarrow \nu \bar\nu \gamma$
at the $Z_1$-pole, in a class of $E_6$ inspired models with a light additional
neutral vector boson \cite{Aytekin}; from the order of
$\mu_{\nu_{\tau}} < O(1.1 \times 10^{-6} \mu_{B})$
Keiichi Akama {\it et al.} derive and apply model-independent bounds on the
anomalous magnetic moments and the electric dipole moments of leptons and quarks
due to new physics \cite{Keiichi}. However, the limits obtained in Ref. \cite{Keiichi}
are for the tau neutrino with an upper bound of $m_\tau < 18.2$ $MeV$ which is
the current experimental limit. It was pointed out in Ref. \cite{Keiichi}
however, that the upper limit on the mass of the electron neutrino and data from
various neutrino oscillation experiments together imply that none of the active
neutrino mass eigenstates is heavier than approximately 3 $eV$. In this case,
the limits given in Ref. \cite{Keiichi} are improved by seven orders of magnitude.
The limit $\mu_{\nu_{\tau}} < 5.4 \times 10^{-7} \mu_{B}$ $(90 \hspace{1mm} \% \hspace{1mm} C.L.)$
is obtained at $q^2=0$ from a beam-dump experiment with assumptions on the $D_s$
production cross section and its branching ratio into $\tau \nu_\tau$ \cite{A.M.Cooper},
thus severely restricting the cosmological annihilation scenario \cite{G.F.Giudice}.
Our results in Tables 1 and 2 for $\phi=-0.009, -0.004, 0, 0.004$,
$\theta_{E_6}=37.8^o$ and $M_{Z_\theta}=7M_{Z_1}$ confirm the bound obtained
by the L3 Collaboration \cite{L3} as well as the bound obtained in Ref. \cite{Aytekin}.

In the case of the electric dipole moment, other upper limits reported in
the literature are \cite{Escribano,Keiichi}:

\begin{eqnarray}
\mid d(\nu_{\tau})\mid &\leq& 5.2 \times 10^{-17} \mbox{$e$ cm} \hspace*{4mm} {\mbox {95 \% C.L.},}\\
\mid d(\nu_{\tau})\mid &<& O(2 \times 10^{-17} \mbox{$e$ cm}).
\end{eqnarray}

Our bounds for the EDM given in Table 2 compare favorably with the
limits given in Eqs. (22, 23). On the other hand, it seems that in
order to improve these limits it might be necessary to study
direct CP-violating effects \cite{M.A.Perez}.

In summary, we conclude that the estimated bounds for the tau neutrino magnetic
and electric dipole moments are almost independent of the experimental
allowed values of the $\phi$ parameter of the LRSM. In the limit
$\phi=0$ and $M_{Z_\theta}\to \infty$, our bound takes the value previously
reported in Ref. \cite{T.M.Gould} for the SM. The bounds in the MM and the
EDM are not affected for the additional neutral vector boson $Z_\theta$ since
its mass is higher than $Z_1$ at $\sqrt{s}=M_{Z_1}$. But at higher center-of-mass
energies $\sqrt{s}\sim M_{Z_\theta}$, the $Z_\theta$ contribution to the cross
section becomes comparable with $Z_1$. In addition, the analytical and numerical
results for the total cross-section have never been reported in the literature
before and could be of some practical use for the scientific community.

\vspace{8mm}

\begin{center}
{\bf Acknowledgments}
\end{center}

We acknowledge support from CONACyT and SNI (M\'exico).

\newpage

\begin{figure}[t]
\begin{center}
\begin{minipage}[h]{120mm}
\epsfig{file=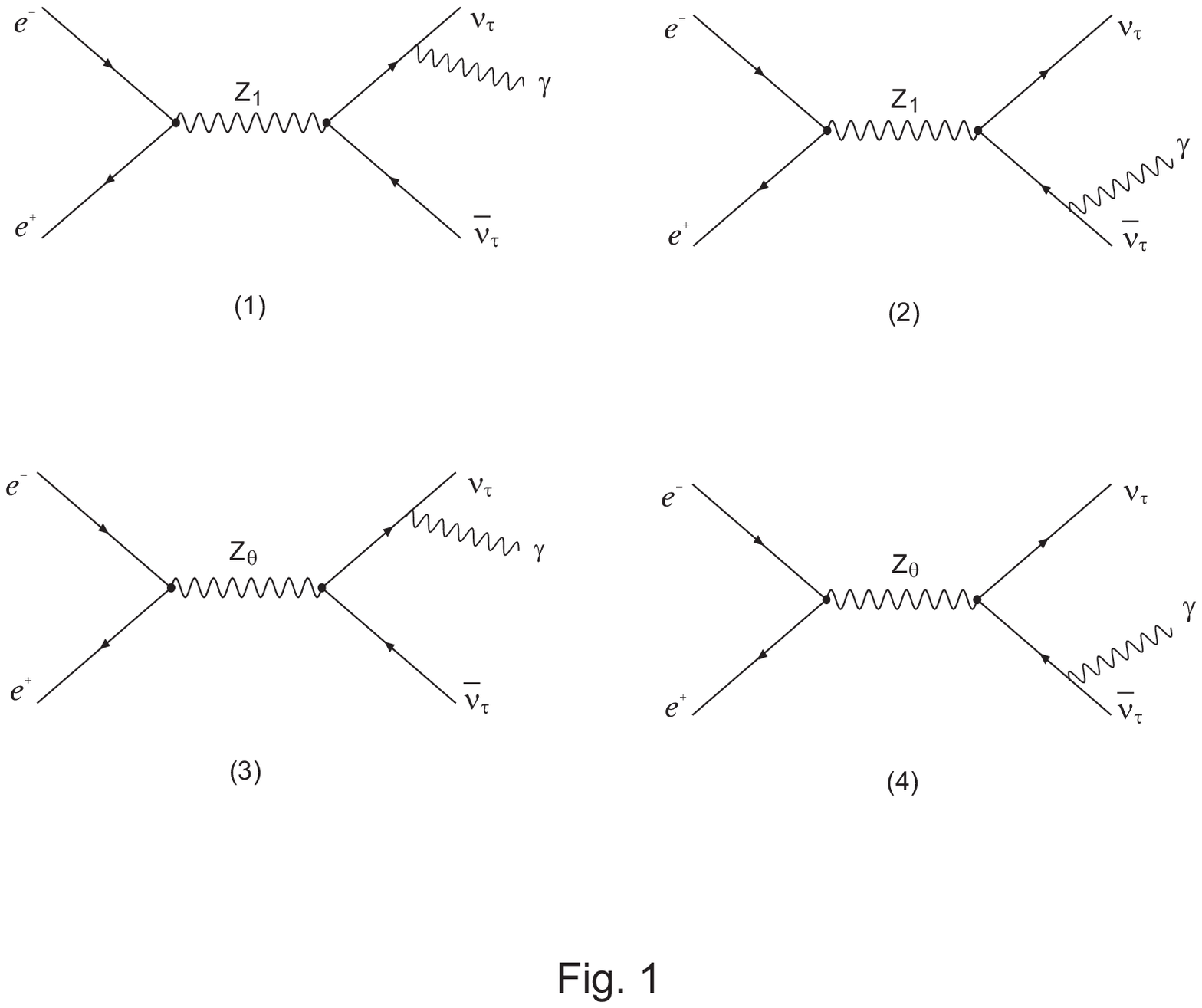, width=130mm}
\end{minipage}
\caption{The Feynman diagrams contributing to the process
$e^{+}e^{-}\rightarrow \nu \bar\nu \gamma$ in a left-right symmetric model,
and in the $E_6$ model.}
\end{center}
\end{figure}

\vspace*{1mm}

\newpage

\begin{figure}[t]
\begin{center}
\begin{minipage}[h]{120mm}
\epsfig{file=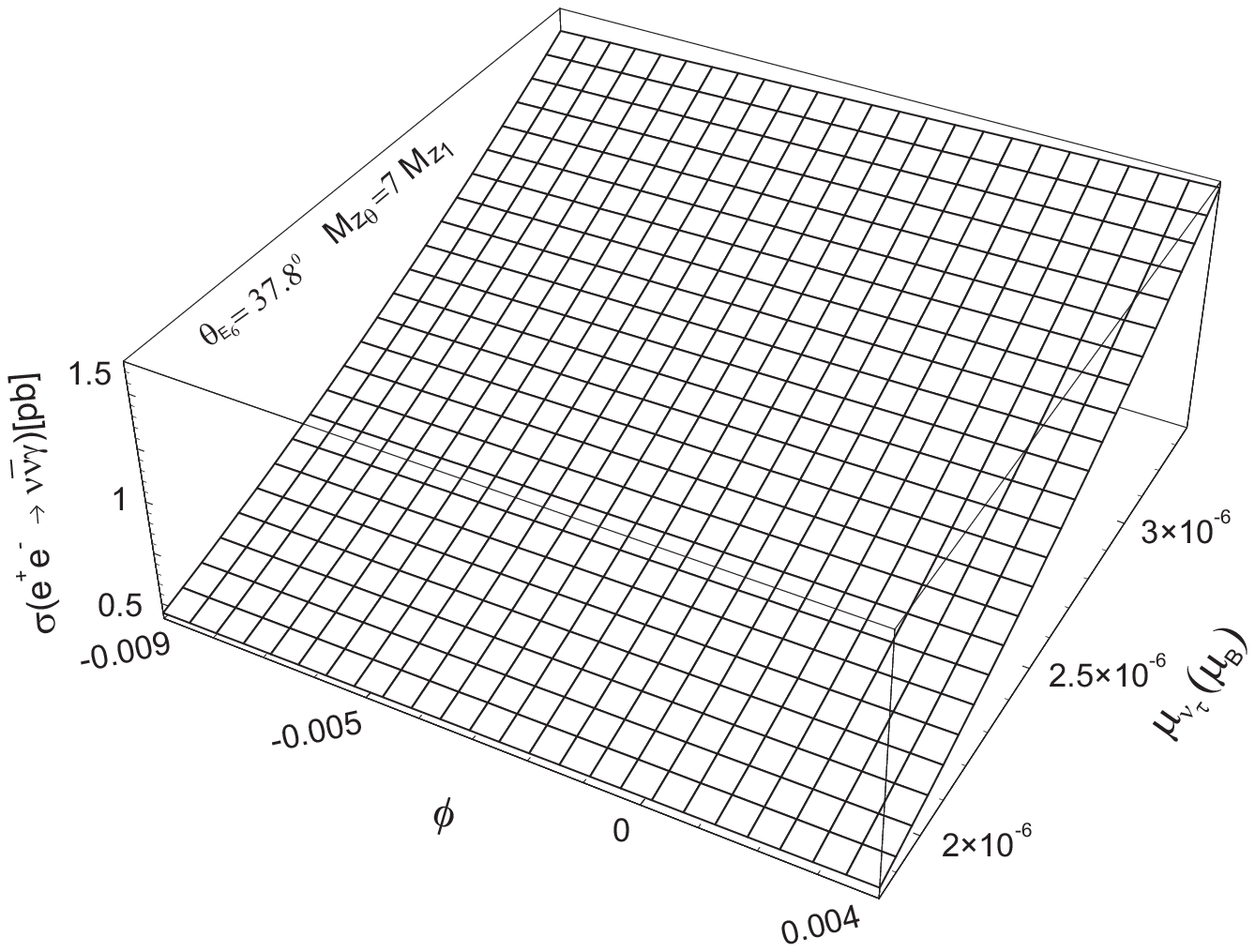, width=130mm}
\end{minipage}
\caption{The total cross section for $e^{+}e^{-}\rightarrow \nu \bar\nu \gamma$
as a function of $\phi$ and $\mu_{\nu_\tau}$ (Tables 1 and 2), with
$\theta_{E_6} = 37.8^o$ and $M_{Z_\theta}=7M_{Z_1}$.}
\end{center}
\end{figure}

\end{document}